%% file: RelevanceJudgementInternals.tex
\documentclass[sigconf,anonymous=false]{acmart}

\usepackage{xcolor}
\usepackage{booktabs}
\usepackage{subcaption}
\usepackage{booktabs}
\usepackage{graphicx}   
\usepackage{xstring}    
\usepackage{booktabs}
\usepackage{multirow}
\usepackage{caption} 
\usepackage{makecell}
\usepackage{tcolorbox}
\usepackage{bm}

\newcommand{\miracl}{\texttt{MIRACL}}
\newcommand{\trecdl}{\texttt{TREC DL20}}

\definecolor{promptframe}{RGB}{110,110,110}
\definecolor{promptback}{RGB}{248,248,248}
\definecolor{prompttitleback}{RGB}{235,235,235}
\definecolor{prompttitletext}{RGB}{40,40,40}

\newcommand{\mycaption}[1]{\caption{{\rm{#1}}}}

\tcbset{
  promptstyle/.style={
    enhanced,
    colback=promptback,
    colframe=promptframe,
    coltitle=prompttitletext,
    colbacktitle=prompttitleback,
    boxrule=0.6pt,
    arc=2mm,
    left=2mm,
    right=2mm,
    top=1.5mm,
    bottom=1.5mm,
    boxsep=1mm,
    fonttitle=\bfseries,
    title filled=true,
    width=\linewidth,
    sharp corners=downhill,
    before skip=8pt,
    after skip=8pt
  }
}

\usepackage[table]{xcolor}
\usepackage{colortbl}

\definecolor{boldgray}{gray}{0.88}

\newcommand{\sidelabel}[2]{%
    \raisebox{#1}{\rotatebox[origin=c]{90}{\textbf{\large #2}}}%
}

\definecolor{VibrantBlue}{rgb}{0.0, 0.2, 1.0}

\AtBeginDocument{%
  }

\setcopyright{acmlicensed}
\copyrightyear{2026}
\acmYear{2026}
\acmDOI{XXXXXXX.XXXXXXX}
\acmConference[CIKM'26]{the 35th ACM International Conference
on Information and Knowledge Management}{November 7–11,
  2026}{Rome, Italy}

\acmISBN{978-1-4503-XXXX-X/2018/06}

\usepackage{enumitem}
\usepackage{tcolorbox}
\tcbuselibrary{skins}

\begin{document}



\title{LLMs Encode Relevance as a Layer-Wise Cross-Lingual Signal}

\author{Pietro Bernardelle} 
\email{p.bernardelle@uq.edu.au}
\orcid{0009-0003-3657-9229}
\affiliation{%
  \institution{The University of Queensland}
  \city{Brisbane}
  \state{}
  \country{Australia}
}

\author{Samaneh Mohtadi} 
\email{s.mohtadi@uq.edu.au}
\orcid{0009-0003-0980-6254}
\affiliation{%
  \institution{The University of Queensland}
  \city{Brisbane}
  \state{}
  \country{Australia}
}

\author{Stefano Civelli} 
\email{s.civelli@uq.edu.au}
\orcid{0009-0003-4982-9565}
\affiliation{%
  \institution{The University of Queensland}
  \city{Brisbane}
  \state{}
  \country{Australia}
}

\author{Joel Mackenzie} 
\email{joel.mackenzie@uq.edu.au}
\orcid{0000-0001-7992-4633}
\affiliation{%
  \institution{The University of Queensland}
  \city{Brisbane}
  \state{}
  \country{Australia}
}

\author{Gianluca Demartini} 
\email{g.demartini@uq.edu.au}
\orcid{0000-0002-7311-3693}
\affiliation{%
  \institution{The University of Queensland}
  \city{Brisbane}
  \state{}
  \country{Australia}
}





\renewcommand{\shortauthors}{Bernardelle et al.}

\begin{abstract}
Large language models (LLMs) are increasingly used in information retrieval (IR) pipelines as relevance judges and re-rankers. 
Yet most analyses remain output-centric, evaluating generated labels or scores while offering limited insight into how relevance is represented inside the model.
In this work, we study whether query--document ($q$--$d$) relevance is linearly decodable from residual-stream activations in instruction-tuned LLMs, how this internal signal compares with generated relevance judgments, and whether it transfers across languages. Using the \trecdl~and \miracl~ evaluation collections, we guide medium-scale LLMs (4--9B parameters) with UMBRELA-style relevance judgment prompts, extract last-token activations from every transformer layer, and train linear probes to predict relevance labels. We compare probe predictions with generated judgments and use \trecdl~ to test whether probe-derived pseudo-labels preserve system rankings when compared with human judgments.
Our results suggest that $q$--$d$ relevance is encoded as a depth-dependent internal signal: probe performance is weak in early layers and strongest in middle-to-late layers, indicating that relevance becomes more linearly accessible after contextual integration. Most importantly, in several models, validation-selected probes match or outperform generated judgments and better preserve system rankings, revealing a separation between internal relevance representation and external expression. Multilingual experiments further suggest partial cross-language portability, although transfer remains weaker than within-language decoding.
Overall, this work provides a representation-level perspective on LLM-based relevance assessment. Layer-wise probing can help diagnose where relevance emerges, when generated judgments fail to reflect internally available evidence, and how relevance representations vary across languages, datasets, and model families.
\end{abstract}



\maketitle

\section{Introduction}
\label{sec:introduction}
Relevance is the central construct around which information retrieval (IR) systems are designed, optimized, and evaluated. Classical retrieval models expose relatively interpretable signals for estimating relevance, including term overlap, inverse document frequency, query likelihood, BM25 scores, and engineered ranking features \cite{manning2008introduction,robertson2009probabilistic,liu2009learning,sparck1972statistical,ponte1998language}. Although these signals are imperfect proxies for relevance, they offer an explanatory basis for understanding why a document is retrieved or ranked highly for a query. This transparency has historically supported both system development and evaluation: retrieval failures could often be traced to missing vocabulary overlap, poor term weighting, insufficient document evidence, or weaknesses in ranking features \cite{furnas1987vocabulary, voorhees2001philosophy}.

This explanatory picture becomes less clear as large language models (LLMs) are increasingly integrated into IR pipelines. LLMs are now used for query expansion, document reranking, answer generation, conversational search, and relevance assessment \cite{zhu2025large,gao2023retrieval,radlinski2017theoretical}. In these settings, relevance is no longer mediated only by sparse lexical signals or explicit ranking features, but also by high-dimensional internal representations formed during the model's forward pass \cite{vaswani2017attention,elhage2021mathematical,geva2021transformer}. The model may output a binary relevance label, a graded judgment, a ranked list, or a natural-language explanation, but the internal computation that gives rise to this output remains opaque. This creates a growing challenge for explainable IR (XIR): as retrieval systems increasingly depend on LLMs, explanations based only on observable inputs and outputs may be insufficient for understanding how relevance judgments are formed. A model may internally encode information that is aligned with the true relevance label while failing to express it correctly in its final output. Conversely, a model may produce apparently accurate labels by exploiting shallow artifacts, such as topical overlap or prompt-specific response patterns. In both cases, output-level evaluation alone cannot distinguish whether relevance is represented internally, where in the model this representation emerges, or how robust it is across prompts, languages, and candidate difficulty.

In this work, we study whether query--document ($q$--$d$) relevance is encoded as a layer-wise linearly decodable signal in instruction-tuned LLMs. While prior work in XIR has begun to analyze internal mechanisms of relevance computation, this literature has largely focused on activation-patching interventions, attention-based analyses, role- or prompt-induced changes in relevance judgment, and the decodability of human-engineered IR features from ranking models~\cite{liu2025large,wang2026role,10.1145/3626772.3657841,chen2026relevance,chowdhury2025probing}. These studies provide important evidence that relevance-related behavior is reflected in model internals. However, it remains unclear whether $q$--$d$ relevance itself, as defined by external relevance judgments, is directly recoverable from the residual stream, where this signal emerges across layers, and whether it is robust across languages.  To address this gap, we organize our work around the following research questions:

\begin{enumerate}[label={\textbf{RQ\arabic*}}, leftmargin=*]
    \item\label{rq:decodability} Is $q$--$d$ relevance linearly decodable from the residual stream of instruction-tuned LLMs?

    \item\label{rq:comparison} How do internal relevance signals compare with explicitly generated relevance judgments?
    
    \item\label{rq:language} How well do internal relevance signals transfer across languages?

\end{enumerate}

We use \texttt{TREC Deep Learning 2020 (DL20)}~\cite{craswell2021overviewtrec2020deep} and \miracl~\cite{zhang-etal-2023-miracl} to guide medium-scale instruction-tuned LLMs with UMBRELA-style relevance-assessment prompts, extract residual-stream activations from the final prompt token at every transformer layer during the same forward pass, and train linear probes to predict human relevance labels. This setup allows us to compare probe-derived judgments with generated outputs, trace relevance decodability across model depth, and evaluate multilingual transfer.
We show that $q$--$d$ relevance is encoded as a layer-wise, linearly decodable signal in the residual stream of instruction-tuned LLMs. This signal is weak in early layers and strongest in middle-to-late layers, and in several settings it shows stronger agreement with human relevance judgments than the model’s generated judgments. We further find that relevance decodability transfers across languages, suggesting a partially cross-lingual internal representation.
Overall, this work contributes to XIR by offering a representation-level perspective on LLM-based relevance judgment. Rather than treating an LLM judge only as a black-box label generator, we examine whether relevance is internally available, where it becomes accessible, and whether it transfers across languages. This provides a basis for more diagnostic, transparent, and reliable use of LLMs in IR evaluation.

\section{Background and Motivation}
\paragraph{LLMs as Relevance Judges.}
The high cost and limited scalability of manual relevance assessment have made LLM-based judging an increasingly prominent direction in IR evaluation \cite{merlo2025cost,rahmani2024llmjudge}. Early work shows that, under carefully designed prompting protocols, LLMs can generate relevance labels that approximate professional assessors and yield system rankings broadly comparable to those obtained from human judgments~\cite{thomas2024large, upadhyay2024llmspatch}. Frameworks such as LLMJudge formalize LLM-based relevance assessment pipelines and enable systematic comparison of judging configurations~\cite{rahmani2025judging}, while large-scale studies using standardized templates, including UMBRELA \cite{upadhyay2024umbrela}, further show that LLM-generated judgments can support scalable evaluation and detailed analyses of agreement patterns and failure modes~\cite{upadhyay2024large}. 

Recent work has moved beyond asking whether LLMs can approximate human labels in aggregate and has instead examined the conditions under which LLM judgments are reliable. 
For example, \citet{keller2026formalized} shows that formalized information needs can improve LLM relevance judgments, suggesting that LLM assessors may require the same kind of topic specification traditionally used in Cranfield-style evaluation. 
Other recent studies examine how LLM judging varies across languages, 
how role-play affects zero-shot LLM rankers~\cite{wang2026role}, how stratified sampling can reduce human effort in validating LLM judgments~\cite{merlo2026reducing}, and how summarization, relevance scales, and query--document representations affect LLM-based assessment quality~\cite{mohtadi2026effect,mohtadi2026query,zamolo2026large}. 

Together, these studies indicate that LLM relevance judgments are not simply interchangeable with human assessments: their reliability depends on prompting, task formulation, language, judgment scale, and validation protocol.
This makes it important to evaluate LLM judges not only at the label level, but also at the system-ranking level. 
Structured assessor disagreement can change system orderings even when overall label agreement appears acceptable~\cite{bailey2008exchangeable,carterette2010assessorerror,gera2024justrank}.

\paragraph{Internal Representations \& Explainability in IR} While the line of work introduced above establishes LLM relevance judging as an output-level evaluation problem, there remains a limited understanding of how these judgments arise and how we can interpret the internal mechanics of LLM judges. 
Specifically, it remains unclear how $q$--$d$ relevance is represented inside the model before a judgment is generated, and whether this representation evolves across layers and transfers across languages.

{\emph{Linear probing}} provides a natural tool for studying this question, as it tests whether information about a target property can be decoded from hidden activations using a simple linear model~\cite{alain2016understanding}. 
In LLMs, probing has been used to analyze how models encode semantic meaning, task-relevant abstractions, and higher-level meta-properties of model behavior~\cite{civelli2026shared,lugoloobi2025llms,zhao2024analysing}. 
In multilingual settings, related work suggests that internal representations often contain a partially shared, language-agnostic structure, particularly in intermediate layers, before later layers specialize toward language-specific generation~\cite{schut2025multilingual,li2025exploring,wendler2024llamas}.

Compared with this broader interpretability literature, XIR has traditionally focused more on observable ranking signals, model attention, feature attribution, or post-hoc explanations than on how relevance is represented internally in LLMs~\cite{anand2022explainable}. 
Recent work has started to close this gap, including activation-patching studies of relevance computation in neural retrieval models and LLMs~\cite{liu2025large,wang2026role,10.1145/3626772.3657841}, layer-wise analyses of internal attention for zero-shot re-ranking~\cite{chen2026relevance}, and probing work showing that activations of ranking LLMs encode human-engineered IR features~\cite{chowdhury2025probing}. 
However, it remains unclear whether human-defined $q$--$d$ relevance itself is linearly decodable from instruction-tuned LLM residual streams, how this signal compares with the model's own generated judgments, and whether it transfers across languages.

Building on the output-level literature on LLM relevance assessment and the representation-level literature on LLM interpretability, we study $q$--$d$ relevance as an internal signal. 
Specifically, we ask whether relevance can be decoded from residual-stream activations, where it is most accessible across model depth, how probe-derived judgments compare with direct LLM inference, and whether the decoded signal exhibits multilingual portability.

\section{Methodology}
\label{sec:methodology}
Our methodology is organized around three complementary experiments ($\S$\ref{sec:experimental-setup}). First, we test whether $q$--$d$ relevance is linearly decodable from residual-stream activations, using \trecdl~passage judgments~\cite{craswell2021overviewtrec2020deep} and the English portion of our sampled \miracl~\cite{zhang-etal-2023-miracl} collection. Second, we compare two sources of relevance information: \emph{explicit output inference}, obtained by parsing the model's generated relevance judgment, and \emph{internal probe inference}, obtained by training linear probes~\cite{alain2016understanding} on layer-wise activations extracted during the same relevance-judgment forward pass on \trecdl. Finally, we return to the full multilingual \miracl~sample to evaluate whether internally decoded relevance transfers across languages, comparing same-language probing with cross-language transfer.

\subsection{Datasets}
\label{sec:data}

\paragraph{TREC DL20}
We select the \trecdl~ passage-ranking task \cite{chowdhury2025probing} as our primary English testbed for studying internal relevance decodability and its impact on downstream IR evaluation. Derived from the MS MARCO passage corpus \cite{bajaj2018msmarcohumangenerated, craswell2021ms}, \trecdl~ provides graded human relevance judgments and officially submitted system runs.\footnote{For convenience, we will use the term \emph{document} throughout the paper to refer both to documents in document-ranking settings and to passages in passage-ranking settings.} This gives us a web-search-style retrieval setting and allows us to evaluate probe-derived relevance estimates both at the judgment level and at the system level (by testing whether pseudo-labels preserve official system rankings). 

\trecdl~ relevance labels are graded as $r \in \{0,1,2,3\}$. We use these judgments in both binary and graded settings. For binary relevance, we consider two standard thresholding strategies. The first uses an inclusive threshold, where the final judgment $y=0$ if $r=0$, and $y=1$ if $r \geq 1$.
This treats any degree of judged relevance as relevant and therefore makes the TREC binary setting more comparable to \miracl's binary labels. 
The second thresholding strategy follows the standard \trecdl~ relevance cutoff used for binary evaluation, where $y=0$ if $r<2$ and $y=1$ if $r \geq 2$.
This treats only clearly relevant documents as positive examples, yielding a more conservative relevance definition. We report both $r \geq 1$ and $r \geq 2$ binarizations to test whether the decodability of relevance is robust to the choice of relevance threshold. For graded experiments, we retain the original $0$--$3$ labels and evaluate ordinal relevance prediction directly.

\paragraph{MIRACL}
While \trecdl~ provides the official system runs needed to evaluate whether probe-derived relevance estimates preserve downstream IR rankings, it does not provide a native multilingual retrieval setting for testing cross-language relevance portability.
We therefore complement it with \miracl~\cite{zhang-etal-2023-miracl}, a large-scale multilingual retrieval benchmark built from language-specific Wikipedia collections and human relevance judgments, as our multilingual testbed. 
The dataset provides native queries and documents across multiple languages, while also providing human judgments that serve as an external gold standard for evaluating both internal probes and generated relevance labels.

From the dataset training split, we sampled $q$--$d$ pairs from seven languages, comprising three higher-resource languages (English, Chinese, and French), together with four languages selected to broaden typological and script diversity (Arabic, Japanese, Swahili, and Telugu). For each language, we sample 250 queries and construct balanced binary $q$--$d$ examples by selecting one relevant and one non-relevant document per query; these documents are sampled from the available \miracl~ judgments using a fixed random seed, yielding the same evaluation examples for all models. Each example consists of a query, a candidate document, and a human relevance label $y \in \{0,1\}$, where $y=1$ denotes relevant and $y=0$ denotes non-relevant.

Because both queries and documents come from the corresponding \miracl~ language collection, our experiments test relevance representations in native multilingual retrieval settings rather than in translated variants of the same English task. This allows us to ask whether relevance is encoded in a language-transferable form or remains tied to language-specific lexical and surface-form features. 

\begin{figure}[t]
    \centering
    \input{miracl-prompt}
    \caption{The prompt used for relevance assessment on the \miracl~ collection.}
    \label{fig:binary-umbrela}
\end{figure}

\begin{table*}[t]
\centering
\small
\mycaption{Parsing success and predicted-label distribution for model-generated UMBRELA judgments, compared with the target human-label distribution. \trecdl~ uses the graded prompt with labels 0--3; \miracl~ uses the binary prompt with labels 0--1 (Figure \ref{fig:binary-umbrela}). Parsed is the percentage of total generations successfully parsed. Label columns report the percentage of successfully parsed outputs assigned to each label, with counts in parentheses. The Target row reports the distribution of the original human relevance labels used for evaluation.}
\label{tab:parse_label_distribution}
\setlength{\tabcolsep}{5pt}
\begin{tabular}{lrrrrrrrr}
\toprule[1.5pt]
\multirow{2}{*}{\textbf{Model}} 
& \multicolumn{5}{c}{\textbf{\trecdl~}} 
& \multicolumn{3}{c}{\textbf{\miracl~}} \\
\cmidrule(lr){2-6} \cmidrule(lr){7-9}
& \textbf{Parsed} & \textbf{0} & \textbf{1} & \textbf{2} & \textbf{3}
& \textbf{Parsed} & \textbf{0} & \textbf{1} \\
\midrule[1.5pt]
\textbf{Target} 
& -- 
& 68.3\% (7780) 
& 17.0\% (1940) 
& 9.0\% (1020) 
& 5.7\% (646)
& -- 
& 50.0\% (1750) 
& 50.0\% (1750) \\

\midrule
Qwen2.5-7B 
& 100.0\% & 66.2\% (7536) & 16.4\% (1869) & 16.3\% (1856) & 1.1\% (125)
& 100.0\% & 57.5\% (2011) & 42.5\% (1489) \\

Qwen3-8B 
& 100.0\% & 36.0\% (4095) & 33.3\% (3795) & 20.6\% (2345) & 10.1\% (1151)
& 100.0\% & 36.1\% (1263) & 63.9\% (2237) \\

Qwen3.5-9B 
& 100.0\% & 37.9\% (4318) & 35.3\% (4018) & 6.1\% (697) & 20.7\% (2353)
& 100.0\% & 41.1\% (1437) & 58.9\% (2063) \\

Llama3.1-8B 
& 96.5\% & 23.5\% (2578) & 24.1\% (2645) & 35.6\% (3915) & 16.9\% (1855)
& 87.8\% & 14.4\% (442) & 85.6\% (2630) \\

Aya-Expanse-8B 
& 95.8\% & 28.0\% (3057) & 13.8\% (1506) & 21.2\% (2312) & 37.0\% (4038)
& 94.5\% & 4.6\% (153) & 95.4\% (3155) \\

Gemma3-4B 
& 100.0\% & 23.2\% (2641) & 23.6\% (2682) & 26.8\% (3048) & 26.5\% (3015)
& 99.6\% & 29.5\% (1029) & 70.5\% (2457) \\
\bottomrule[1.5pt]
\end{tabular}
\end{table*}

\subsection{Models}
\label{sec:models}

We evaluate instruction-tuned LLMs \cite{ouyang2022training} at a comparable medium scale (4--9B parameters), focusing on models that are large enough to perform relevance judgment but small enough to allow manageable experimental runtimes. Our model suite includes models from the Qwen, Llama, Aya, and Gemma families, including: Qwen2.5-7B-Instruct \cite{qwen2025qwen25technicalreport}, Qwen3-8B \cite{yang2025qwen3}, Qwen3.5-9B \cite{qwen3.5}, Llama-3.1-8B-Instruct \cite{grattafiori2024llama}, Aya-Expanse-8B \cite{dang2024ayaexpansecombiningresearch}, Gemma-3-4B-it \cite{gemmateam2025gemma3technicalreport}. Both Qwen3 and Qwen3.5 have been run using the {\emph{no-thinking}} inference configuration.

Our model selection serves three purposes. First, it keeps model size within a narrow range, so differences in probe performance are less likely to be driven by scale. Second, it includes multiple independently trained model families, allowing us to test whether layer-wise relevance decodability is model-specific or appears across architectures and training pipelines. Third, it includes models with different degrees of multilingual specialization. The Qwen models additionally provide a controlled family-level comparison across versions.

\subsection{Prompting and Output Inference}
\label{sec:prompting}

For both datasets, each $q$–-$d$ pair is formatted using an UMBRELA-style relevance-judgment prompt containing the query and document text. For \trecdl~ we use the standard graded UMBRELA format \cite{upadhyay2024umbrela}, whereas for \miracl~ we construct a binary adaptation of the UMBRELA-style prompt, replacing the graded relevance scale with labels 0 and 1 to match the dataset’s binary judgments (see Figure \ref{fig:binary-umbrela}).  We apply each model's native chat template before inference, ensuring that the prompt is presented in the instruction-following format expected by that model. 
Judgments are generated using deterministic decoding --- with the temperature set to $0$ --- so that output-inference results are not affected by sampling variability.

The generated output is parsed to extract the final relevance label (either binary or graded). 
We report, for each model and dataset, the percentage of generations that were successfully parsed and the resulting predicted-label distribution (see Table~\ref{tab:parse_label_distribution}). These statistics are reported separately from agreement metrics because output-inference baselines depend not only on the model's relevance assessment, but also on whether the model follows the requested answer format. Probe-based inference does not have this limitation. 
















\subsection{Activation Extraction}
\label{sec:activation-extraction}
For each model, prompt, and $q$--$d$ input, we run a forward pass and extract residual-stream activations from every transformer layer. Let $h^{(\ell)}_{i} \in \mathbb{R}^{d}$ denote the residual-stream activation for example $i$ at layer $\ell$, where $d$ is the model hidden dimension. We extract the activation at the final input token, immediately before answer generation. This position has attended to the full prompt, including the query, document, and relevance-judgment instruction, and therefore represents the model state from which the final relevance judgment is generated.

For a model with $L$ layers, this produces a set of layer-wise representations:
\[
\left\{h^{(1)}_{i}, h^{(2)}_{i}, \ldots, h^{(L)}_{i}\right\}.
\]
We train probes independently at each layer, yielding a full depth profile of relevance decodability.

\subsection{Linear Probes}
\label{sec:probing}

We use linear probes, following ridge regularization~\cite{hoerl1970ridge}, to test whether relevance is linearly recoverable from residual-stream activations. For each model layer $\ell$, a probe maps the activation $h_i^{(\ell)}$ for $q$--$d$ pair $i$ to a relevance prediction:
\[
\hat{y}_{i}^{(\ell)} = f_{\ell}(h_i^{(\ell)}).
\]
For binary relevance, $y_i \in \{0,1\}$, we train regularized logistic-regression probes. For graded \trecdl~ experiments, where $y_i \in \{0,1,2,3\}$, we use ordinal-logistic probes rather than treating the four relevance levels as unordered classes. 
We deliberately use linear probes for diagnostic clarity rather than maximum predictive performance. Success under this setting indicates that relevance is available in a simple, linearly accessible form. Future work could examine whether more expressive probing methods recover additional relevance structure.

\paragraph{Grouped data splits.}
We split examples by query rather than by individual $q$--$d$ pair. This prevents leakage across train, validation, and test sets: all documents associated with the same query are assigned to the same split. Without this grouping, documents for a query could appear in training while other documents for the same query appear in testing, allowing probes to exploit query-specific information rather than testing generalisation to unseen queries.

For \trecdl, we use grouped 5-fold cross-validation by query. In our setup, \trecdl~ contains 54 queries, which we partition into five outer folds. Four folds contain 11 held-out test queries, and one fold contains 10 held-out test queries. For each outer fold, the remaining queries are used for model development and are further divided into training and validation groups. Probes are fitted on the training groups, while the validation groups are used to select the regularisation strength and probing layer. Final performance is then computed on the held-out test queries for that fold. In our experiments, we report the mean performance across the five outer folds and use the standard deviation to summarize fold-level variability.

For \miracl, we use a fixed grouped split in which 60\% of query-language groups are used for training, 20\% for validation, and the remaining 20\% for testing. The grouping key combines document language and query id, so that each query within a given language appears in only one split. We reuse this split for both same-language probing and cross-language transfer. In same-language probing, probes are trained and evaluated within the same language. In cross-language transfer, probes trained on a source language are evaluated on held-out test examples from target languages.

\paragraph{Hyperparameter selection.}
During the training phase of each probe, we sweep the regularization strength over the same logarithmic grid for all models and layers:
\[
\alpha \in \{10^{-4}, 3{\cdot}10^{-4}, 10^{-3}, 3{\cdot}10^{-3}, 10^{-2}, 3{\cdot}10^{-2}, 0.1, 0.3, 1, 3, 10, 30, 100\}.
\]
At each layer, we select the value of $\alpha$ that maximizes validation agreement with human labels. For graded \trecdl~ probes, validation selection uses quadratic weighted Cohen's $\kappa$ \cite{cohen1968weighted}, reflecting the ordinal structure of the relevance labels. For binarized \trecdl~ and \miracl~ probes, $\alpha$ is selected via standard Cohen's $\kappa$ \cite{cohen1960coefficient}.

\paragraph{Layer selection.}
We distinguish between two layer-selection settings. In the main validation-selected setting, both the regularization strength $\alpha$ and the probing layer are selected using validation performance computed on held-out development labels. The selected probe is then evaluated once on test data that was not used for either training or model selection. This corresponds to the system-level setting used for our main results -- performance reflects what would be obtained when choosing a layer and hyperparameter from labeled development data and then applying that choice to unseen $q$--$d$ pairs.

We also report an oracle setting, where the best layer is selected retrospectively using test performance. This is not a valid model-selection procedure because it uses information from the held-out test set. We include it only as an upper-bound diagnostic to show how much relevance is linearly recoverable somewhere in the residual stream, independent of whether that layer can be selected reliably from validation data. Oracle results are therefore not used to support the main claims.

\subsection{Experimental Settings}
\label{sec:experimental-setup}

We organize the experiments around the three research questions introduced earlier ($\S$\ref{sec:introduction}). Rather than treating \trecdl~ and \miracl~ as separate analyses, we use them as complementary settings for studying the same underlying question: whether relevance is internally represented in LLM activations, how this internal signal compares with explicit model judgments, and whether it transfers across languages.

\paragraph{Experiment 1: Layer-wise relevance decodability.}
To address \ref{rq:decodability}, we first evaluate whether $q$--$d$ relevance is linearly recoverable from residual-stream activations. We use \trecdl~ together with the English subset of our sampled \miracl~ $q$--$d$ pairs, allowing us to test decodability in both a standard English IR test collection with graded relevance judgments, and a native retrieval benchmark with binary labels. For each model and dataset, we train probes independently at each layer and evaluate how predictive performance changes across model depth. 

For \trecdl, we evaluate both binary and graded formulations: binary probes use the $r \geq 1$ and $r \geq 2$ relevance thresholds, while graded probes retain the original $0$--$3$ relevance labels. For \miracl, probes are trained to predict binary relevance labels. This setting provides the main layer-wise analysis of where relevance becomes most accessible inside the model.

\paragraph{Experiment 2: Internal probing versus generated judgments.}
To address \ref{rq:comparison}, we compare the relevance signal recovered from internal activations with the relevance judgments explicitly generated by the same models on \trecdl. The output-inference baseline is obtained by parsing each model's deterministic UMBRELA-style judgment, while the probing setting uses validation-selected linear probes trained on the corresponding layer-wise activations. 

We compare generated and probe-derived labels at two levels: agreement with human $q$--$d$ relevance judgments, and preservation of gold-judgment system rankings (i.e., system rankings induced by the official human judgments over official submitted runs). This tests whether internally decoded relevance supports both individual relevance decisions and downstream IR evaluation.

\paragraph{Experiment 3: Multilingual relevance portability.}
To address \ref{rq:language}, we extend the \miracl~ analysis beyond English and evaluate whether the internal relevance signal transfers across languages. We use the sampled \miracl~ $q$--$d$ pairs from all selected languages and keep the same query-safe train, validation, and test splits described above. 

We consider two multilingual probing settings. In the same-language setting, probes are trained and evaluated within the same language, measuring how strongly relevance is decodable in each native language collection. In the cross-language setting, probes trained on one source language are evaluated on held-out examples from other target languages:
\[
M_{s,t} = \mathrm{Perf}(f_s, D_t),
\]
where $f_s$ denotes a probe trained on source language $s$ and $D_t$ denotes the held-out test split for target language $t$. Comparing these two settings lets us distinguish language-specific relevance decodability from cross-language portability, while preserving the same layer-wise probing framework used in the earlier experiments.

All experiments were run on a single NVIDIA RTX 5090 GPU. The full experimental pipeline, including deterministic output inference, layer-wise activation extraction, probe training, and hyperparameter selection across layers, required approximately one and a half days of wall-clock time.\footnote{
Code is publicly available at \url{https://anonymous.4open.science/r/probing-relevance/}}

\begin{figure*}[t]
    \centering
    \setlength{\fboxsep}{0pt}      
    \setlength{\fboxrule}{0.4pt}   

    \begin{subfigure}{\linewidth}
        \centering
        \begin{tabular}{@{}c@{\hspace{0.8em}}c@{}}
            \sidelabel{11.8em}{\trecdl} &
            \fbox{%
                \includegraphics[width=0.93\linewidth]{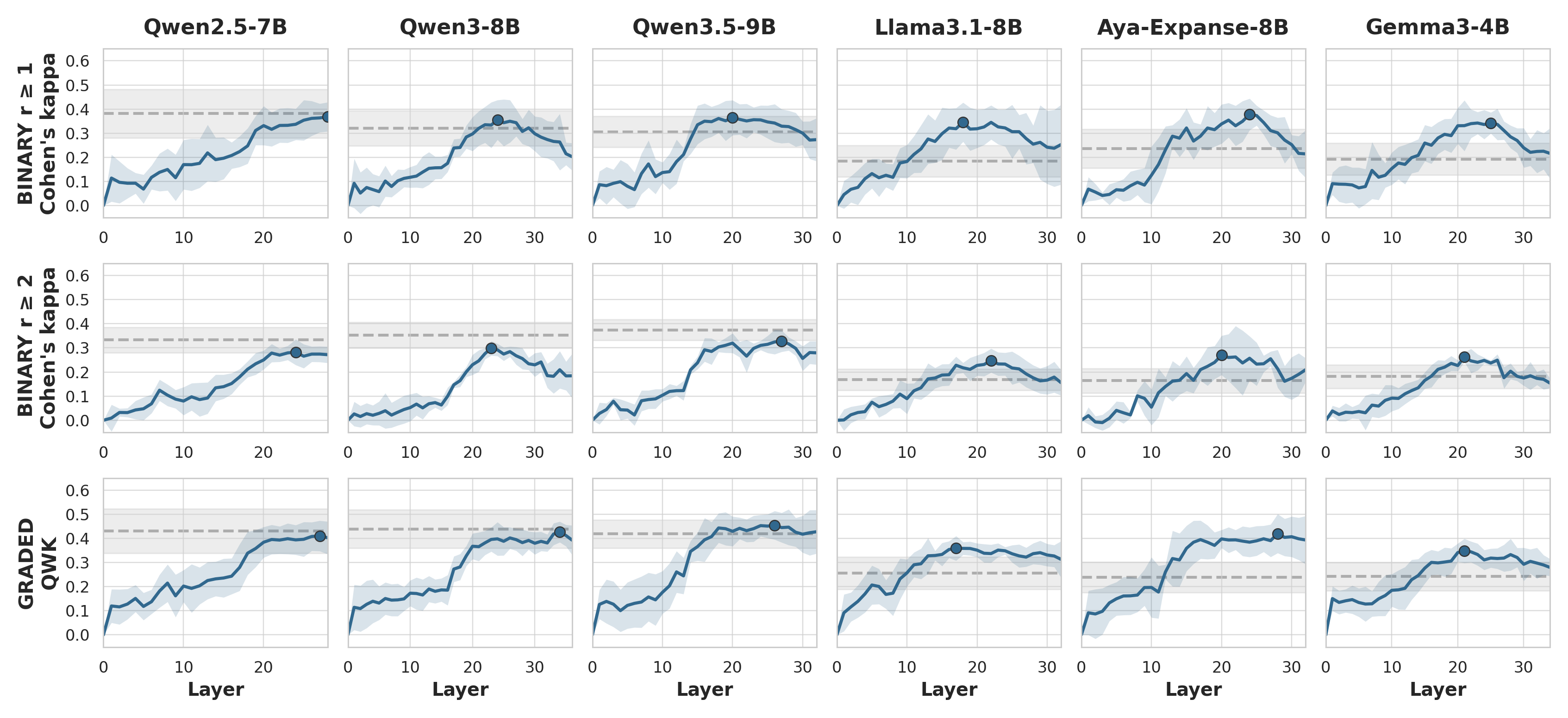}%
            }
        \end{tabular}
        \label{fig:layer-wise-kappa-trec}
    \end{subfigure}

    \vspace{0.5em}

    \begin{subfigure}{\linewidth}
        \centering
        \begin{tabular}{@{}c@{\hspace{0.8em}}c@{}}
            \sidelabel{4.8em}{\miracl~ (en)} &
            \fbox{%
                \includegraphics[width=0.93\linewidth]{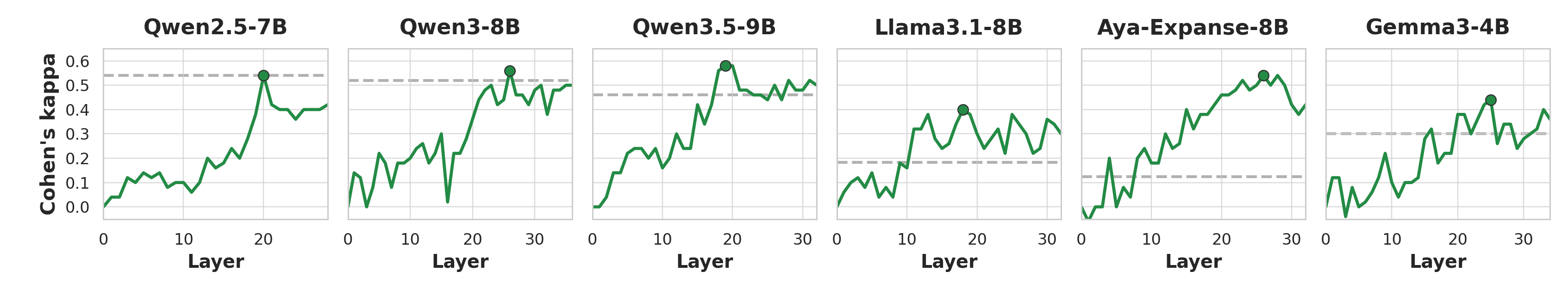}%
            }
        \end{tabular}
        \label{fig:layer-wise-kappa-miracl}
    \end{subfigure}

    \mycaption{Layer-wise probing performance on \trecdl~ and \miracl. \trecdl~ results use grouped query splits and report binary relevance prediction with Cohen's $\kappa$ after binarizing labels, alongside graded relevance prediction with quadratic weighted $\kappa$. For \miracl, we report the English subset results using Cohen's $\kappa$. Lines show held-out probe performance, shaded bands indicate variation across folds. Markers identify the best-performing layers under the corresponding held-out metric. Note that models have different numbers of layers.}
    \label{fig:layer-wise-kappa}
\end{figure*}

\subsection{Evaluation Metrics}
\label{sec:metrics}
We evaluate performance at two levels: label-level agreement and system-ranking stability. 

For \trecdl~ binary experiments, Cohen's $\kappa$ is the primary label-level agreement metric~\cite{cohen1960coefficient}. Cohen's $\kappa$ measures agreement between predicted and human relevance labels while accounting for chance agreement. We use standard Cohen's $\kappa$ for the binarised \trecdl~ settings, and quadratic weighted Cohen's $\kappa$ for the graded setting over the original $0$--$3$ labels~\cite{cohen1968weighted}. Quadratic weighting penalises larger disagreements more heavily than smaller ones, making it appropriate for ordinal relevance labels.

For \trecdl, we additionally evaluate whether generated or probe-derived pseudo-labels preserve official system rankings. In the binarised setting, systems are ranked using average precision (AP)\footnote{Also know as ``Mean Average Precision'' (MAP).} computed under both $r \geq 1$ and $r \geq 2$ relevance thresholds~\cite{manning2008introduction}, as well as by Rank-Biased Precision (RBP) with $\phi = 0.9$~\cite{moffat2008rank}, which provides a more explicitly top-weighted view of effectiveness under a persistent-user browsing model. In the graded setting, following standard practice in TREC Deep Learning evaluations, retrieval effectiveness is measured using nDCG@10~\cite{jarvelin2002cumulated}, which supports graded relevance judgments and emphasizes performance at top ranks~\cite{craswell2021overviewtrec2020deep}.
To quantify ranking stability, we compare the system rankings induced by human labels, generated pseudo-labels, and probe-derived pseudo-labels. We use Kendall's $\tau$ as the primary rank-correlation metric because it measures pairwise ordering consistency and is widely used to study sensitivity to assessor variation~\cite{kendall1938new,bailey2008exchangeable,carterette2010assessorerror,gera2024justrank}. We additionally report Spearman's $\rho$ to capture monotonic agreement between system rankings~\cite{spearman1987proof}, and Rank-Biased Overlap (RBO) with $\phi = 0.9$ to emphasize stability among top-ranked systems~\cite{webber2010similarity}.

For \miracl~ binary same-language and cross-language experiments, we also use Cohen's $\kappa$ as the primary label-level agreement metric~\cite{cohen1960coefficient}. Since the \miracl~ labels are binary, $\kappa$ directly measures agreement between predicted and human relevance labels while accounting for chance agreement, making it consistent with the binarised \trecdl~ evaluation.

\section{Results}
\label{sec:results}
In this section, we present the results following the three experimental settings introduced earlier ($\S$\ref{sec:experimental-setup}). We first examine whether relevance labels are linearly recoverable from residual-stream activations -- and where this signal is most accessible across layers (\ref{rq:decodability}). We then compare internally decoded relevance with the judgments explicitly generated by the same models, both at the $q$--$d$ label level and at the system-ranking level (\ref{rq:comparison}). Finally, we evaluate whether the internal relevance signal transfers across languages in native multilingual \miracl~ settings (\ref{rq:language}).

\subsection{Layer-wise Relevance Decodability}
\label{sec:results-rq1}

\paragraph{Relevance decodability increases with model depth.}
Figure~\ref{fig:layer-wise-kappa} shows the layer-wise probing results on \trecdl~ and \miracl~ English. Across models and relevance formulations, probe performance is weakest in early layers and generally increases through the middle of the network, with the strongest values typically appearing in middle-to-late or late layers. 
This provides direct evidence that $q$--$d$ relevance is linearly recoverable from residual-stream activations, but it is not uniformly available across layer depths (\ref{rq:decodability}). Early representations contain limited relevance information under a linear readout, whereas later representations make human relevance labels more separable. A natural interpretation is that relevance becomes more accessible after the model has integrated the query, document, and relevance-judgment instruction \cite{nijasure2025relevance}. 

This depth profile is consistent with the broader view that transformer representations become increasingly task-structured across layers, and with probing work showing that high-level semantic or task-relevant properties can become linearly accessible in hidden states~\cite{alain2016understanding,lugoloobi2025llms}. In the IR setting, it also aligns with recent mechanistic studies suggesting that relevance judgment is a progressive computation rather than a single final-layer decision~\cite{liu2025large,chen2026relevance,nijasure2025relevance,chowdhury2025probing}. Our results complement this literature by showing that externally defined $q$--$d$ relevance itself, rather than only relevance-related mechanisms or engineered IR features, is recoverable from the residual stream and becomes more linearly accessible across depth.

\begin{figure}
    \centering
    \includegraphics[width=1\linewidth]{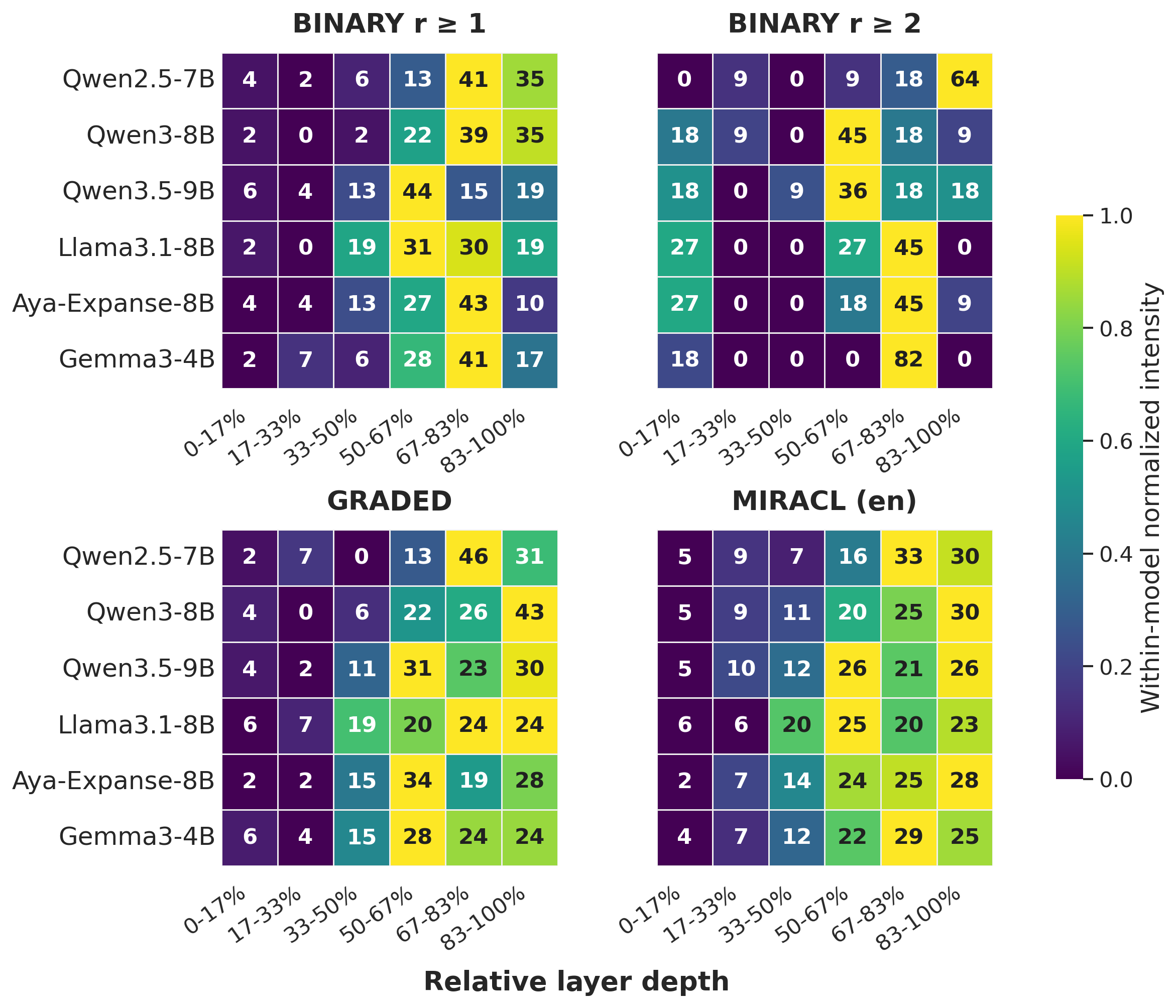}
    \mycaption{Distribution of query-level probe informativeness across relative model depth on \trecdl~ and \miracl~ (en). Each panel shows one probing setup: binary probes are evaluated with Cohen’s $\kappa$, while graded probes are evaluated with quadratic weighted $\kappa$. Rows correspond to models and columns to relative-depth bins. Cell values report the percentage of fractional Top-1 query mass assigned to each bin, splitting ties evenly between equally best-performing bins. Color intensity is normalized within each model row, highlighting where the most informative layers are concentrated for that model under each labeling setup.}
    \label{fig:layer_depth}
\end{figure}

\paragraph{The depth pattern holds at the query level.}
While Figure~\ref{fig:layer-wise-kappa} shows that relevance decodability improves with depth on average, it does not show whether this pattern is driven by a small number of examples or whether it holds more broadly across queries.
Figure~\ref{fig:layer_depth} complements this view by asking where individual queries achieve their strongest probe performance as a function of relative model depth. Across \trecdl~ binary and graded relevance, and \miracl~ English subset, the earliest depth bins rarely contain the largest share of query-level Top-1 mass. Instead, informative layers concentrate in the $50$--$67\%$, $67$--$83\%$, and $83$--$100\%$ depth ranges.
The query-level results strengthen the interpretation that relevance becomes most linearly accessible after substantial model computation, rather than being concentrated in early prompt- or lexical-processing layers (\ref{rq:decodability}).

Interestingly, the stricter $r \geq 2$ setting is more diffuse than $r \geq 1$, suggesting that answer-bearing relevance may require a broader range of middle-to-late computations. In this setting, grade $1$ passages, which are related to the query but not sufficient to answer it, are grouped with grade $0$ passages as negatives, while grades $2$ and $3$ form the positive class~\cite{craswell2021overviewtrec2020deep}. The probe must therefore distinguish answer-bearing relevance from a negative class that mixes fully irrelevant passages with merely related ones. This may cause different queries to reach their strongest linear separability at different depths, with some peaking earlier and others requiring middle-to-late representations. The overall concentration of Top-1 mass beyond the earliest bins nevertheless supports the broader conclusion that relevance is generally most accessible after substantial model computation.

\begin{table}[t]
    \centering
    \small
    \setlength{\tabcolsep}{5pt}
    \mycaption{Grouped 5-fold \trecdl~ test-set agreement for binary and graded relevance. Binary probes use Cohen's $\kappa$, while graded probes use ordinal-logistic quadratic weighted $\kappa$ (QWK). System rows select layers by the corresponding validation metric. Oracle rows select layers by the corresponding test metric and are included only as an upper-bound reference for probe performance. Values are mean $\pm$ standard deviation across outer folds. \underline{Underlined} values indicate the better result between the deployable system probe and the inference output baseline for the same model and metric; oracle values are not considered for highlighting.}
    \label{tab:trec_dl20_grouped_binary_ordinal_logistic_test_kappa}
    \begin{tabular}{clccc}
      \toprule[1.5pt]
      &
      \textbf{Model} &
      {\renewcommand{\arraystretch}{0.75}\begin{tabular}{c}\textbf{BINARY}\\\bm{$r\ge1$ $\kappa$}\end{tabular}} &
      {\renewcommand{\arraystretch}{0.75}\begin{tabular}{c}\textbf{BINARY}\\\bm{$r\ge2$ $\kappa$}\end{tabular}} &
      {\renewcommand{\arraystretch}{0.75}\begin{tabular}{c}\textbf{GRADED}\\\textbf{QWK}\end{tabular}} \\
      \midrule[1.5pt]
    & & \multicolumn{3}{c}{\textit{\textbf{System: layer selected by validation metric}}} \\
    \multirow{12.5}{*}{\rotatebox[origin=c]{90}{\textbf{PROBE}}}
    & Qwen2.5-7B  & 0.354 $\pm$ 0.083 & 0.272 $\pm$ 0.034 & 0.395 $\pm$ 0.061 \\
    & Qwen3-8B  & \underline{0.344 $\pm$ 0.097} & 0.298 $\pm$ 0.030 & 0.427 $\pm$ 0.043 \\
    & Qwen3.5-9B & \underline{0.364 $\pm$ 0.073} & 0.324 $\pm$ 0.051 & \underline{0.443 $\pm$ 0.061} \\
    & Llama3.1-8B & \underline{0.346 $\pm$ 0.081} & \underline{0.248 $\pm$ 0.050} & \underline{0.322 $\pm$ 0.039} \\
    & Aya-Expanse-8B & \underline{0.379 $\pm$ 0.065} & \underline{0.237 $\pm$ 0.134} & \underline{0.393 $\pm$ 0.053} \\
    & Gemma3-4B  & \underline{0.331 $\pm$ 0.105} & \underline{0.237 $\pm$ 0.011} & \underline{0.321 $\pm$ 0.040} \\
    \cmidrule{2-5}
    & & \multicolumn{3}{c}{\textit{\textbf{Oracle: layer selected by test metric}}} \\
    & Qwen2.5-7B  & 0.368 $\pm$ 0.060 & 0.282 $\pm$ 0.053 & 0.410 $\pm$ 0.063 \\
    & Qwen3-8B  & 0.355 $\pm$ 0.082 & 0.298 $\pm$ 0.030 & 0.427 $\pm$ 0.043 \\
    & Qwen3.5-9B & 0.364 $\pm$ 0.073 & 0.328 $\pm$ 0.044 & 0.454 $\pm$ 0.059 \\
    & Llama3.1-8B & 0.346 $\pm$ 0.081 & 0.248 $\pm$ 0.050 & 0.359 $\pm$ 0.051 \\
    & Aya-Expanse-8B & 0.379 $\pm$ 0.065 & 0.269 $\pm$ 0.081 & 0.420 $\pm$ 0.081 \\
    & Gemma3-4B & 0.343 $\pm$ 0.044 & 0.262 $\pm$ 0.020 & 0.349 $\pm$ 0.050 \\
    \midrule
    & & \multicolumn{3}{c}{\textit{\textbf{Inference output baseline}}} \\
    \multirow{5}{*}{\rotatebox[origin=c]{90}{\textbf{INFERENCE}}}
    & Qwen2.5-7B  & \underline{0.381 $\pm$ 0.101} & \underline{0.333 $\pm$ 0.052} & \underline{0.430 $\pm$ 0.092} \\
    & Qwen3-8B  & 0.321 $\pm$ 0.073 & \underline{0.353 $\pm$ 0.054} & \underline{0.439 $\pm$ 0.079} \\
    & Qwen3.5-9B  & 0.306 $\pm$ 0.064 & \underline{0.375 $\pm$ 0.043} & 0.419 $\pm$ 0.058 \\
    & Llama3.1-8B  & 0.185 $\pm$ 0.065 & 0.168 $\pm$ 0.045 & 0.255 $\pm$ 0.067 \\
    & Aya-Expanse-8B  & 0.241 $\pm$ 0.081 & 0.165 $\pm$ 0.051 & 0.239 $\pm$ 0.063 \\
    & Gemma3-4B  & 0.193 $\pm$ 0.066 & 0.182 $\pm$ 0.049 & 0.242 $\pm$ 0.059 \\
    \bottomrule[1.5pt]
    \end{tabular}
\end{table}

\begin{table*}[t]
\centering
\setlength{\tabcolsep}{4pt}
\newcommand{\innersep}{\hspace{0.4\dimexpr2\tabcolsep\relax}}
\mycaption{Leaderboard agreement between official \trecdl~ systems under gold labels and pseudo-label rankings induced by UMBRELA outputs (inference) or validation-selected probes. For binary relevance, systems are ranked using either AP or RBP ${\phi=0.9}$ under two binarization thresholds, $r \geq 1$ and $r \geq 2$. For graded relevance, systems are ranked using nDCG@10 with the original TREC grades retained, using the ordinal-logistic graded probe for validation-selected probe rows. Agreement with the official leaderboard is measured using RBO ${\phi=0.9}$, Spearman's $\rho$, and Kendall's $\tau$. \underline{Underline} indicates the higher value between probe and inference for each metric on each model.}
\label{tab:trec_dl20_system_leaderboard_reformatted}
\begin{tabular}{ll
  c@{\innersep}c@{\innersep}c
  c@{\innersep}c@{\innersep}c
  c@{\innersep}c@{\innersep}c
  c@{\innersep}c@{\innersep}c
  c@{\innersep}c@{\innersep}c}
\toprule[1.5pt]
\textbf{Model} & \textbf{Type}
& \multicolumn{6}{c}{\textbf{BINARY} \bm{$r \geq 1$}}
& \multicolumn{6}{c}{\textbf{BINARY} \bm{$r \geq 2$}}
& \multicolumn{3}{c}{\textbf{GRADED}} \\
\cmidrule(lr){3-8}
\cmidrule(lr){9-14}
\cmidrule(lr){15-17}
& & \multicolumn{3}{c}{\textbf{AP}}
  & \multicolumn{3}{c}{\textbf{RBP} $\phi = 0.9$}
  & \multicolumn{3}{c}{\textbf{AP}}
  & \multicolumn{3}{c}{\textbf{RBP$\phi = 0.9$}}
  & \multicolumn{3}{c}{\textbf{nDCG@10}} \\
\cmidrule(lr){3-5}\cmidrule(lr){6-8}
\cmidrule(lr){9-11}\cmidrule(lr){12-14}
\cmidrule(lr){15-17}
& & \textbf{RBO} & \bm{$\rho$} & \bm{$\tau$}
  & \textbf{RBO} & \bm{$\rho$} & \bm{$\tau$}
  & \textbf{RBO} & \bm{$\rho$} & \bm{$\tau$}
  & \textbf{RBO} & \bm{$\rho$} & \bm{$\tau$}
  & \textbf{RBO} & \bm{$\rho$} & \bm{$\tau$} \\
\midrule[1.5pt]

\multirow{2}{*}{Qwen2.5-7B}
& Probe
& 0.847 & \underline{0.970} & \underline{0.878}
& 0.714          & \underline{0.969} & \underline{0.875}
& 0.651          & 0.958          & 0.850
& \underline{0.782} & \underline{0.979} & \underline{0.885}
& 0.626          & \underline{0.986} & \underline{0.912} \\
& Inference
& \underline{0.868} & 0.968          & 0.861
& \underline{0.768} & 0.966          & 0.862
& \underline{0.816} & \underline{0.971} & \underline{0.870}
& 0.781          & 0.975          & 0.882
& \underline{0.628} & 0.978          & 0.885 \\
\midrule

\multirow{2}{*}{Qwen3-8B}
& Probe
& \underline{0.858} & \underline{0.989} & \underline{0.917}
& \underline{0.836} & \underline{0.979} & \underline{0.902}
& \underline{0.720} & \underline{0.966} & \underline{0.849}
& 0.714          & \underline{0.990} & \underline{0.918}
& 0.636          & \underline{0.987} & \underline{0.916} \\
& Inference
& 0.736          & 0.623          & 0.486
& 0.698          & 0.941          & 0.813
& 0.688          & 0.931          & 0.790
& \underline{0.789} & 0.979          & 0.891
& \underline{0.650} & 0.983          & 0.906 \\
\midrule

\multirow{2}{*}{Qwen3.5-9B}
& Probe
& \underline{0.876} & \underline{0.978} & \underline{0.891}
& \underline{0.832} & \underline{0.982} & \underline{0.897}
& \underline{0.731} & \underline{0.955} & 0.826
& 0.822          & 0.983          & 0.906
& 0.668          & 0.986          & 0.911 \\
& Inference
& 0.719          & 0.600          & 0.472
& 0.788          & 0.935          & 0.806
& 0.716          & 0.948          & \underline{0.829}
& \underline{0.824} & \underline{0.989} & \underline{0.922}
& \underline{0.711} & \underline{0.989} & \underline{0.926} \\
\midrule

\multirow{2}{*}{Llama3.1-8B}
& Probe
& \underline{0.878} & \underline{0.948} & \underline{0.829}
& 0.714          & \underline{0.973} & \underline{0.881}
& \underline{0.733} & \underline{0.918} & \underline{0.752}
& \underline{0.769} & \underline{0.965} & \underline{0.848}
& 0.419          & 0.962          & 0.857 \\
& Inference
& 0.696          & 0.584          & 0.452
& \underline{0.725} & 0.878          & 0.737
& 0.535          & 0.553          & 0.425
& 0.679          & 0.924          & 0.773
& \underline{0.472} & \underline{0.977} & \underline{0.884} \\
\midrule

\multirow{2}{*}{Aya-Expanse-8B}
& Probe
& \underline{0.892} & \underline{0.986} & \underline{0.914}
& 0.758          & \underline{0.977} & \underline{0.877}
& \underline{0.620} & \underline{0.951} & \underline{0.834}
& \underline{0.740} & \underline{0.964} & \underline{0.871}
& 0.644          & \underline{0.983} & \underline{0.901} \\
& Inference
& 0.655          & 0.707          & 0.557
& \underline{0.878} & 0.939          & 0.818
& 0.548          & 0.695          & 0.536
& 0.713          & 0.951          & 0.829
& \underline{0.683} & 0.969          & 0.868 \\
\midrule

\multirow{2}{*}{Gemma3-4B}
& Probe
& \underline{0.872} & \underline{0.958} & \underline{0.855}
& \underline{0.698} & \underline{0.978} & \underline{0.887}
& \underline{0.529} & \underline{0.956} & \underline{0.838}
& \underline{0.673} & \underline{0.975} & \underline{0.871}
& \underline{0.675} & \underline{0.976} & \underline{0.883} \\
& Inference
& 0.685          & 0.570          & 0.452
& 0.633          & 0.657          & 0.496
& 0.503          & 0.532          & 0.414
& 0.663          & 0.933          & 0.788
& 0.452          & 0.959          & 0.836 \\

\bottomrule[1.5pt]
\end{tabular}
\end{table*}

\subsection{Internal Probing vs. Generated Judgments}
\label{sec:results-rq2}
Having established that $q$--$d$ relevance is linearly decodable from residual-stream activations, we next ask how this internal signal compares with respect to the model's explicitly generated judgment (\ref{rq:comparison}). This comparison is important since output-level LLM judging does not isolate relevance assessment alone. A generated label also depends on whether the model follows the provided instruction, uses the relevance scale in a calibrated way, and produces an answer in the requested format. Prior work has shown that LLM-generated judgments can approximate human assessments and preserve system rankings under controlled prompting protocols~\cite{thomas2024large,upadhyay2024umbrela,upadhyay2024large}. At the same time, judgment quality remains sensitive to prompt design, task formulation, role specification, and relevance criteria~\cite{rahmani2025judging,wang2026role}. We therefore use the probe--output comparison to test whether some of this variability reflects a mismatch between what the model internally represents and what it finally expresses.

\paragraph{Generated judgments and internal probes expose different signals.}
Table~\ref{tab:trec_dl20_grouped_binary_ordinal_logistic_test_kappa} compares validation-selected probes (top grouping) with explicit output inference (bottom grouping) by measuring their agreement with human \trecdl~ labels at the $q$--$d$ label level. The comparison shows that the relative strength of probes and generated judgments varies across models and appears alongside clear differences in the distribution of generated labels (Table \ref{tab:parse_label_distribution}). This is most evident for Qwen2.5-7B, whose generated TREC labels most closely track the target distribution in Table~\ref{tab:parse_label_distribution}, and whose inference output baseline is strongest across all three relevance formulations in Table~\ref{tab:trec_dl20_grouped_binary_ordinal_logistic_test_kappa} compared to validation-selected probes. Qwen3-8B and Qwen3.5-9B are less distributionally aligned than Qwen2.5-7B, but still produce strong output baselines in the stricter $r \geq 2$ and graded settings. This suggests that generated judgments can sometimes express relevance distinctions that are not fully matched by a single-layer linear readout.

The pattern differs for models whose generated labels are strongly skewed toward higher relevance grades. Table~\ref{tab:parse_label_distribution} shows that Llama3.1-8B, Aya-Expanse-8B, and Gemma3-4B assign substantially more mass to labels 2 and 3 than the target label distribution, with Aya-Expanse-8B the most extreme case: 37.0\% of parsed TREC outputs are assigned grade 3, compared with 5.7\% in the target labels. For these models, validation-selected probes achieve higher agreement with human labels than the generated outputs in both binary settings as well as the graded setting. Similar probe gains also appear for Qwen3-8B and Qwen3.5-9B under $r \geq 1$. 

These results suggest that weak generated judgments do not necessarily imply that relevance information is absent from the model. Rather, in several models, a single-layer linear readout from the residual stream recovers a relevance signal that is more aligned with human judgments than the model's final generated label. One possible implication is that the final generated judgment is not simply a direct expression of the model's internal relevance estimate. Instead, the generation step may transform, recalibrate, or distort an internal relevance signal according to model-specific response biases, label-scale preferences, or instruction-following behavior. Under this hypothesis, probing exposes a representation that can be closer to the target relevance criterion than the discrete label ultimately produced by the model (\ref{rq:comparison}).

\paragraph{System probes are close to oracle probes.}
The comparison above treats validation-selected probes as the deployable probing results. Since probe performance varies across layers, we also ask how much performance is lost by selecting layers using validation labels rather than retrospectively choosing the best test layer. To quantify this gap, the oracle rows in Table~\ref{tab:trec_dl20_grouped_binary_ordinal_logistic_test_kappa} (middle grouping) select the layer with the highest test performance, providing an upper-bound diagnostic for probe performance rather than a deployable model-selection procedure.
In most cases, validation-selected probes are close to this oracle upper bound, suggesting that the probe-vs-output comparison is not an artefact of retrospective layer selection; held-out validation labels are usually sufficient to identify a high-performing depth region.

\paragraph{Probe-derived pseudo-labels often preserve system rankings.}
Table~\ref{tab:trec_dl20_system_leaderboard_reformatted} compares the leaderboard induced by gold labels with leaderboards induced by probe-derived pseudo-labels and model-generated relevance judgments. This system-level comparison asks whether the label-level differences we have observed also translate into different rankings of retrieval systems. 

Overall, the results show that probe-derived pseudo-labels often recover the structure of the gold judgment leaderboard, particularly for models whose generated judgments diverge from human labels. This pattern is clearest for Llama3.1-8B, Aya-Expanse-8B, and Gemma3-4B. For these models, leaderboards induced by generated judgments often depart substantially from the official ranking, whereas probe-derived pseudo-labels yield markedly higher agreement. Qwen2.5-7B is a partial exception: its generated judgments already align strongly with the gold judgment leaderboard, leaving less room for probes to improve. Qwen3-8B and Qwen3.5-9B show a more threshold-dependent pattern, with probes performing better under the more permissive $r \geq 1$ setting and generated judgments retaining stronger agreement in some stricter or graded settings.

Across metrics, the probe advantage is most consistent for Spearman's $\rho$ and Kendall's $\tau$, indicating that probes better preserve the global and pairwise ordering of systems. RBO shows a more mixed pattern, suggesting that generated judgments can sometimes better match the very top of the leaderboard, especially under stricter or graded relevance definitions. This variation across metrics is closely related to how relevance is defined for evaluation. When relevance is binarised with the more permissive $r \geq 1$ cut, probe-derived pseudo-labels provide the clearest gains. Under the stricter $r \geq 2$ cut and in the graded nDCG@10 setting, the comparison becomes closer, suggesting that the relative benefit of probes depends partly on how graded relevance judgments are transformed or preserved. This sensitivity is consistent with prior work on relevance-scale transformation, which shows that different mappings from graded to binary judgments can yield distinct agreement patterns and IR system rankings~\cite {han2019transforming}. Overall, these results suggest that internal probes are particularly effective at preserving the broader leaderboard structure, even when generated judgments sometimes better match the top-ranked systems under stricter or graded evaluation settings (\ref{rq:comparison}).

\begin{figure*}[t]
    \centering
    \includegraphics[width=1\linewidth]{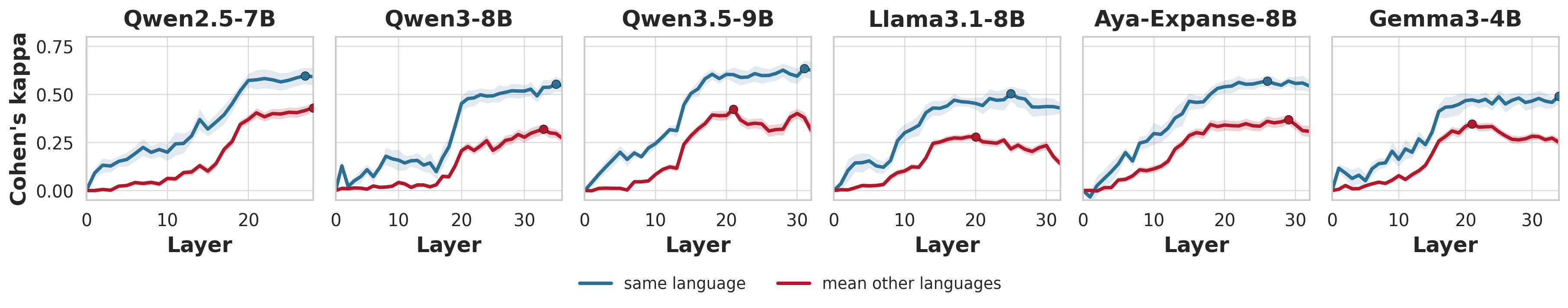}
    \mycaption{Layer-wise portability of binary UMBRELA relevance probes on native \miracl. The blue curve shows the mean same-language held-out Cohen's $\kappa$ across languages, while the red curve shows the mean cross-language Cohen's $\kappa$ when probes are applied to held-out examples from other languages. Shaded bands indicate the standard error of the mean across language pairs. Dots mark the layer with the highest mean Cohen's $\kappa$ for each curve.}
    \label{fig:miracl-all}
\end{figure*}

\begin{figure}[t]
    \centering
    \includegraphics[width=1\linewidth]{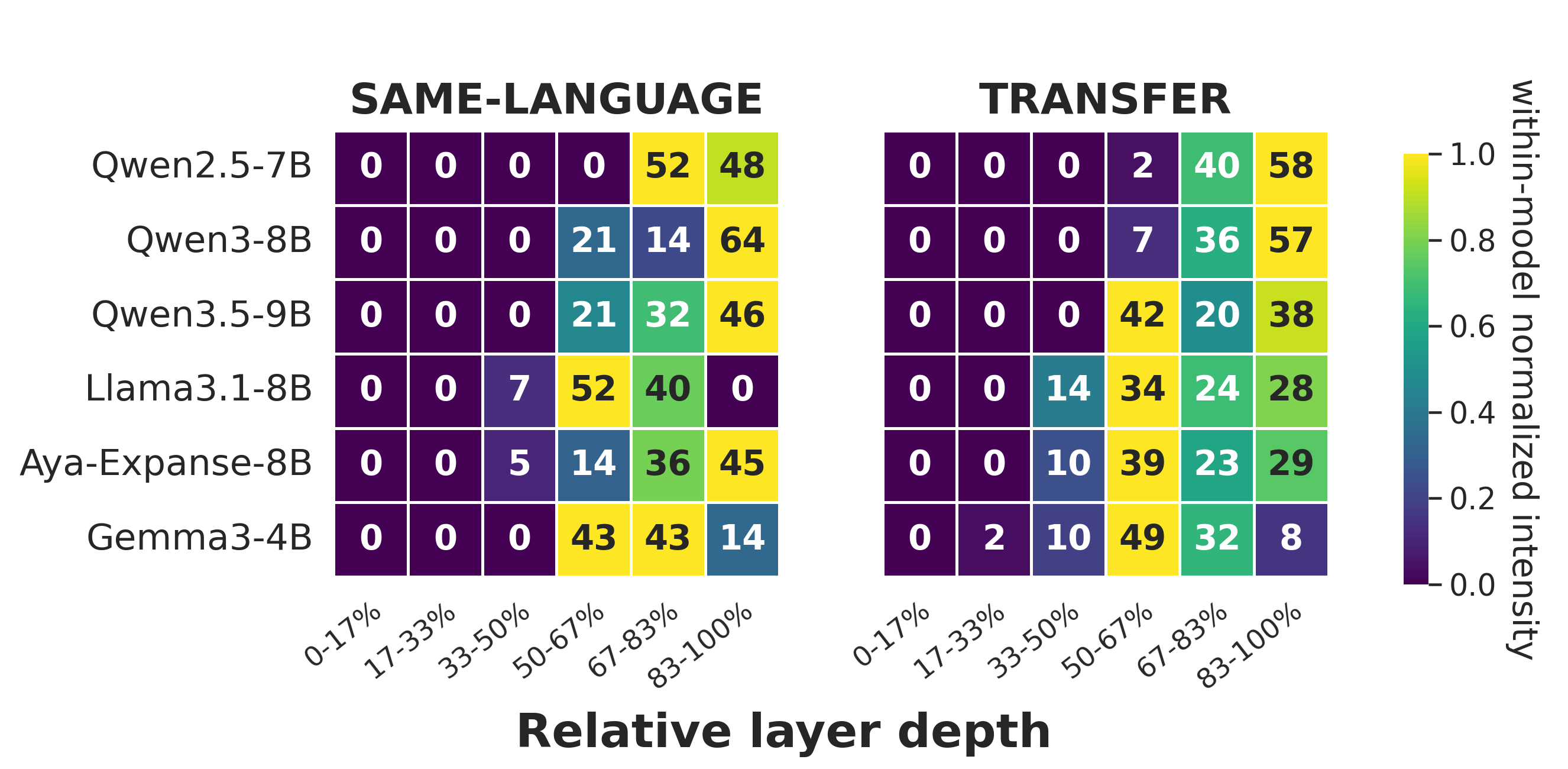}
    \mycaption{Distribution of peak probe portability across relative model depth on native \miracl. Each cell reports the percentage of language-pair evaluations whose highest held-out Cohen's $\kappa$ occurs in the corresponding relative-depth bin, splitting ties evenly across equally best-performing layers. The same-language panel summarizes probes trained and tested within the same language, while the cross-language transfer panel summarizes probes trained on one language and evaluated on held-out examples from another language. Color intensity is normalized within each model row, highlighting where each model’s strongest same-language and transferable relevance signals are concentrated.}
    \label{fig:miracl_peak_portability}
\end{figure}

\subsection{Multilingual Relevance Portability}
\label{sec:results-rq3}

Finally, we ask whether relevance is encoded in a cross-lingual form (\ref{rq:language}). 
Since relevance is a meta-property of the relation between a query and a candidate document, a shared semantic representation should allow a probe trained in one language to recover at least some relevance information in another.
We evaluate \ref{rq:language} on native \miracl~ by comparing same-language probes with cross-language transfer probes across the selected languages and models.

\paragraph{Cross-language probes recover a weaker relevance signal.}
Figure~\ref{fig:miracl-all} compares same-language \miracl~ probes with cross-language transfer probes. Probes trained and evaluated within the same language generally outperform probes transferred across languages, indicating that the recovered relevance signal is not fully language-agnostic. Nevertheless, cross-language performance is not flat. Across models, transfer curves typically improve with depth and reach their strongest values in middle-to-late or late layers, broadly mirroring the depth profile observed in the same-language setting. This suggests a subtle form of portability in which probes trained in one language can recover a non-trivial relevance signal from held-out examples in another (\ref{rq:language}).

This interpretation is consistent with broader multilingual representation studies showing that LLMs can form partially shared semantic spaces across languages while still retaining language-specific structure that affects downstream behavior and output generation~\cite{wendler2024llamas,schut2025multilingual,li2025exploring}.

\paragraph{Transfer peaks are late but less localized than same-language peaks.}
While Figure~\ref{fig:miracl-all} shows that cross-language transfer improves with depth on average, it does not show whether the strongest transfer layers are concentrated in the same depth regions as same-language probes. Figure~\ref{fig:miracl_peak_portability} addresses this by comparing where same-language and cross-language probes reach their strongest performance across relative model depth. 
Same-language peaks are strongly concentrated in late layers, and for several models the largest mass falls in the final $83$--$100\%$ depth bin. Cross-language peaks also tend to occur in middle-to-late and late layers, but they are less sharply localized. For several models, the strongest transfer performance is spread across the $50$--$67\%$, $67$--$83\%$, and $83$--$100\%$ bins rather than collapsing onto the final layers alone. Thus, portability follows the same broad depth profile as same-language relevance decoding, but with a more diffuse peak structure.

This pattern connects the multilingual transfer results to the earlier layer-wise findings ($\S$\ref{sec:results-rq1} and $\S$\ref{sec:results-rq2}). In those experiments, relevance became most linearly accessible only after the model had processed a substantial portion of the network, suggesting that the signal depends on contextual integration of the query, document, and relevance-assessment instruction rather than on shallow lexical features alone. The cross-language results follow this same general trajectory. If transfer were driven primarily by surface-level overlap or prompt-format cues, we would expect the strongest cross-language signal to appear much earlier. Instead, the transferable signal emerges in the same middle-to-late depth region in which same-language relevance becomes accessible, supporting the view that cross-language portability is tied to higher-level semantic representations.


At the same time, the weaker localization of the transfer peaks indicates that this shared representation is only partial. Same-language probes can exploit language-specific and task-specific information that becomes especially pronounced in later layers, whereas cross-language probes must rely on components of the representation that remain aligned across languages. This helps explain why transfer improves with depth but remains less stable than same-language decoding. The result is also consistent with work on multilingual LLM internals, which suggests that semantic alignment across languages is often most visible in intermediate or middle-to-late layers, before later computations become more specialized toward language-specific generation~\cite{wendler2024llamas,civelli2026shared}. Our findings extend this view to relevance assessment: the layers that support transfer appear to encode enough shared semantic structure to make relevance partly recoverable across languages, while the more diffuse peaks show that language-specific specialization still limits full portability.

\section{Conclusions}
\label{sec:conclusion}

In this work, we examined LLM-based relevance assessment from a representation-level perspective. Rather than evaluating LLM judges only through their generated labels, we asked whether human-defined $q$--$d$ relevance is encoded in the residual stream, where this signal emerges across layers, how it relates to explicit output judgments, and whether it transfers across languages.

Across \trecdl~\cite{craswell2021overviewtrec2020deep} and \miracl~\cite{zhang-etal-2023-miracl}, our results show that relevance is linearly decodable from internal activations. This signal is weak in early layers and becomes most accessible in middle-to-late and late layers. The pattern is consistent across model families, relevance formulations, and evaluation settings, suggesting that relevance emerges as a depth-dependent internal signal. We also find that internal relevance and generated judgments are related but not equivalent. In some settings, model outputs provide strong relevance estimates; in others, probes recover relevance information that is not faithfully expressed in the final generated label. This points to a representation--expression gap: LLMs may encode relevance-relevant evidence internally even when their explicit judgments are poorly calibrated, skewed toward particular labels, or misaligned with the requested output format. At the system-evaluation level, probe-derived pseudo-labels can preserve \trecdl~ rankings more faithfully than generated outputs, indicating that internal relevance representations are not only useful for diagnosing individual $q$--$d$ decisions, but can also support downstream IR evaluation. Our multilingual results further show that relevance representations are partially portable across languages. Cross-language probes often decode relevance above early-layer baselines and follow a similar depth-dependent trajectory to same-language probes. However, native-language probing remains stronger, indicating that internal relevance representations combine shared multilingual structure with language-specific information. This suggests a promising role for internal signals in multilingual and lower-resource IR, while also cautioning against treating cross-language transfer as a substitute for native-language supervision.

While the scope of this work is limited to UMBRELA-style relevance prompts, medium-scale instruction-tuned LLMs, final-token residual-stream activations, and linear probes, we believe our findings can open several directions for future work. Future studies could extend this analysis to larger and retrieval-specialized models, examine richer activation summaries over query and document spans, and test more expressive probing architectures, such as multilayer perceptrons, while preserving the diagnostic distinction between what is linearly recoverable and what is recoverable through stronger nonlinear decoders.

Overall, our findings suggest that LLM judges should not be treated only as black-box generators of relevance labels. Their internal states provide diagnostic evidence about whether relevance is represented, where it becomes accessible, and when generated judgments fail to reflect internally available information. This opens a path toward more transparent LLM-based IR evaluation, where internal signals can complement output-level judgments and help identify model-, layer-, language-, and task-specific failure modes.

\section*{GenAI Usage Disclosure}

Generative AI tools were used to support the refinement of writing, flow, and clarity in this manuscript. All research design, experiments, analyses, results, interpretations, and substantive scholarly contributions were conducted and produced by the authors.

\balance
\bibliographystyle{ACM-Reference-Format}
\bibliography{references}

\end{document}

%% file: miracl-prompt.tex
\begin{tcolorbox}[promptstyle]

Given a query and a passage, you must provide a binary relevance score with the following meanings:

0 = represents that the passage is not relevant to the query,\\
1 = represents that the passage is relevant to the query.
\medskip

Important Instruction: Assign 0 if the passage is unrelated, only tangentially related, or related to the topic without answering the query. Assign 1 only if the passage contains answer-bearing information for the query.

\medskip
Query: \textbf{\{query\}}\\
Passage: \textbf{\{passage\}}

\medskip
Split this problem into steps:\\
Consider the underlying intent of the search.
Measure whether the passage contains information that satisfies the query intent (M).\\
Measure how trustworthy the passage is (T).\\
Consider the aspects above and decide on a final binary relevance score (O).\\ 
Final score must be an integer value only.
Do not provide any code in the result. Provide the score in the format: \#\#final score: score without providing any reasoning.
\end{tcolorbox}